\documentclass[global,twocolumn]{svjour}
\usepackage{graphics}
\usepackage{epsfig}

\usepackage{amsmath}
\usepackage{amsopn}
\usepackage{amsfonts}
\usepackage{amssymb}
\usepackage{lineno}
\journalname{Appl.Phys.B}

\begin{document}
% %%%%%%%%%%%%%%%% comment for nolinennumbers%%%%%%%%%%%%%%%
\pagewiselinenumbers
%%%%%%%%%%%%%%%%%%%%%%%%%%%%%%%%%%%%%%%%%%%%%%%%%%%%%%%%%%%%
%
%------------------------Title----------------------------------------------------------------------
%
%
%\title{}
\title{Optimized production of a cesium Bose-Einstein condensate}
\author{Tobias Kraemer$^1$, Jens Herbig$^1$, Michael Mark$^1$, Tino Weber$^1$, Cheng
Chin$^1$, Hanns-Christoph N\"agerl$^1$\thanks{corresponding
author}, and Rudolf Grimm$^{1,2}$ }
 \authorrunning{T. Kraemer et al.}
 \institute{Institut f\"ur
Experimentalphysik, University of Innsbruck, A-6020 Innsbruck,
Austria
}                     % Do not remove
\institute{$^1$Institut f\"ur Experimentalphysik, Universit\"at
Innsbruck, Technikerstra{\ss}e 25, A-6020 Innsbruck, Austria\\
$^2$Institut f\"ur Quantenoptik und Quanteninformation,
{\"O}sterreichische Akademie der Wissenschaften, A-6020 Innsbruck,
Austria %\and the second here
}
\email{christoph.naegerl@uibk.ac.at}
\date{Received: date / Revised version: date}

\maketitle
%--------------------Abstract----------------------------------------------------------
%
\begin{abstract}
We report on the optimized production of a Bose-Einstein
condensate of cesium atoms using an optical trapping approach.
Based on an improved trap loading and evaporation scheme we obtain
more than $10^5$ atoms in the condensed phase. To test the
tunability of the interaction in the condensate we study the
expansion of the condensate as a function of scattering length. We
further excite strong oscillations of the trapped condensate by
rapidly varying the interaction strength.
\end{abstract}
%
%
%--------------Introduction------------------------------------------------
%
\section{Introduction}\label{Intro}

Much of the present work in the field of quantum gases relies on
optical trapping techniques and on the ability to tune atomic
interactions. Optical approaches have been recently employed in
several atomic Bose-Einstein condensation experiments
\cite{Barrett2001,Takasu2003,Cennini2003,Weber2003a,Rychtar2004}
and in experiments on the production of ultracold molecular
samples \cite{Chin2003a,Regal2003,Herbig2003,Duerr2004,Xu2003} and
on molecular Bose-Einstein condensates
\cite{Jochim2003,Greiner2003}. The major advantages in optical
traps are the possibility to trap atoms in any sublevel of the
electronic ground state and the ease to adjust the interaction
strength using magnetically induced Feshbach resonances.

The cesium atom is very attractive for experiments with tunable
atomic interactions. The lowest internal quantum state of Cs
%($F=3$,$m_F=3$)
features a unique combination of wide and narrow Feshbach
resonances which are easily accessible at low magnetic fields
\cite{Chin2003}. This results in a great flexibility for tuning
the atomic scattering properties. In particular, magnetic tuning
of the interaction strength has recently allowed the first
realization of a Bose-Einstein condensate (BEC) with Cs atoms
\cite{Weber2003a} and the realization of a two-dimensional
condensate very close to a dielectric surface \cite{Rychtar2004}.
The tunability of the atomic interaction can be exploited in
experiments where one might wish to adjust or to dynamically
change the mean-field interaction of the condensate. Also, the
Feshbach resonances can be used to produce molecules from an
atomic BEC \cite{Donley2002,Herbig2003,Duerr2004,Xu2003} and to
study the transition from an atomic BEC to a molecular BEC. In
this context, a quantum phase transition with an essentially
topological character has been predicted
\cite{Rad2004,Romans2004}. For such and many other intriguing
experiments it is desirable to have a large BEC of Cs atoms as a
starting point.

In this paper we report on the optimized production of an
essentially pure Cs BEC in the lowest internal quantum state with
more than $ 10^5 $ atoms. Since this state cannot be trapped by
purely magnetic means, the path to condensation relies on a
sequence of optical traps. We discuss the loading and transfer
from one trap to the next and give a detailed description of the
evaporation path and of the resulting condensate. As a
demonstration for tunability we measure the expansion energy as a
function of scattering length in time-of-flight experiments. In
particular, we show the ultra-slow expansion of the condensate
after release from the trap for nearly vanishing scattering
length. The release energy corresponds to $ \sim 50 $ pK. Finally,
we present first results when the scattering length is suddenly
stepped and the condensate then starts to oscillate freely in the
trap.

\section{Cesium scattering properties and Feshbach resonances}\label{Csscattering}

Early experiments \cite{Soding1998,Arlt1998} towards condensation
of cesium focused on samples in magnetic traps polarized either in
the upper hyperfine ground state $F=4$, magnetic sublevel $m_F=4$,
or in the lower hyperfine state $F=3,m_F=-3$. Here, $F$ denotes
the total angular momentum and $m_F$ the magnetic quantum number.
The spin relaxation rates were measured to be several orders of
magnitude higher than expected
\cite{GueryO1998a,GueryO1998b,Hopkins2000}. It was later
understood that this is caused by the dipolar relaxation process
induced by the second-order spin-orbit interaction \cite{Leo1998}.
The maximum phase-space density in a small sample of Cs atoms was
a factor of about four away from condensation \cite{Thomas2003}.

The problem of the strong inelastic two-body losses can be
overcome by using the lowest internal state of cesium, $F=3,m_F=3$
\cite{Perrin1998,Vuletic1999,Hammes2000,Han2001}. In this state,
all inelastic two-body processes are endothermic and are thus
fully suppressed at sufficiently low temperature. This state
requires optical trapping since it cannot be captured in a
magnetic trap. Optically trapped atoms can only be efficiently
evaporated by lowering the total potential depth. This process
weakens the confinement of the trapped sample and thus makes it
difficult to achieve sufficiently high elastic collision rates for
effective evaporation. Hence, adjustability of the collisional
properties is very helpful for a fast evaporation strategy.

\begin{figure}[t]
\begin{center}
\epsfig{file=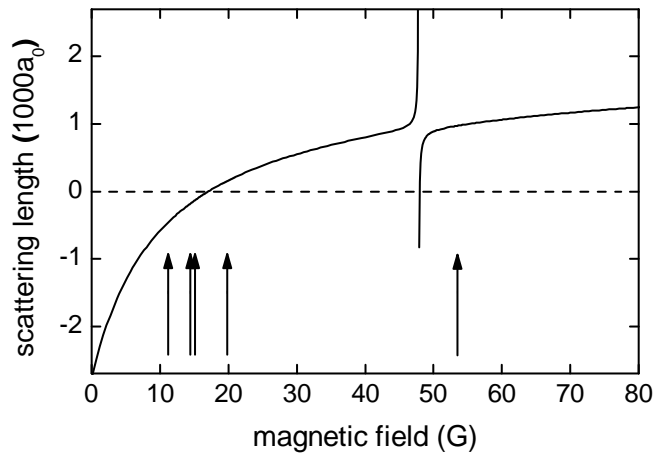,width=1.0\linewidth} \caption{ Scattering
length as a function of magnetic field for the state $F=3, m_F=3$.
There is a relatively broad Feshbach resonance at $ 48.0 $ G due
to coupling to a d-wave molecular state. The arrows indicate
several very narrow resonances at $ 11.0, 14.4, 15.0, 19.9 $ and $
53.5 $ G, which result from coupling to g-wave molecular states.
The data is taken from \cite{Chin2003}. \label{Csscatteringlength}
}
\end{center}
\end{figure}

The success in condensing Cs \cite{Weber2003a} largely relies on
the fact that the s-wave scattering length for the $F=3,m_F=3$
state can be tuned to moderate and positive values by the
application of relatively low dc magnetic fields \cite{Chin2003}.
As Fig.~\ref{Csscatteringlength} shows, an external magnetic field
allows for precise tuning of the atomic scattering length $a$ from
negative to positive values. Positive scattering lengths in the
range between zero and one thousand $ a_0 $ are attained for
magnetic fields of a few ten Gauss; $ a_0 $ denotes Bohr's radius.
In particular, there is a gentle zero-crossing of the scattering
length near $ 17 $ G \cite{Vuletic1999}. Here, the interaction of
atoms in a BEC is effectively switched off. Several narrow
higher-order Feshbach resonances \cite{Chin2003}, caused by
coupling to d- and g-wave molecular states, enable very rapid
control of the atomic scattering properties. With the magnetic
field being a free parameter in our optical trapping approach, we
can take full advantage of this tunability of the s-wave
scattering length.

For Cs in the $F=3, m_F=3$ ground state the process of three-body
recombination is the dominant loss and heating mechanism
\cite{Weber2003b}. In a recombination process, three atoms
collide, two of them form a molecule, and the third atom takes
away two thirds of the binding energy according to energy and
momentum conservation. The atoms that form the molecule are
usually lost, and the third atom is either lost or it deposits its
share of the binding energy in the sample. Heating of the sample
is the combination of ``anti-evaporation'' and recombination
heating \cite{Weber2003b}. To a good approximation, the three-body
recombination rate scales with the fourth power of the scattering
length. Unfortunately, the prefactor in this scaling law is
measured to be relatively large \cite{Weber2003b}. To minimize
this heating, the recombination products should be removed quickly
from the trap. It is thus important to assure that the sample is
not operated too deep in the hydrodynamic regime and that the
evaporation is efficient in {\it all} directions.  Arbitrarily
increasing the scattering length to speed up the forced
evaporation is therefore not possible without sacrificing cooling
efficiency. Within these limits, tuning the scattering length
allows for an optimization of the evaporation for given trap
parameters. For example, for the low initial densities in a large
reservoir trap the evaporation may be sped up by increasing the
scattering length. In a later trapping stage with a higher atomic
density the scattering length should be reduced to optimize the
ratio of good to bad collisions.

\section{BEC production}\label{BECprod}

\subsection{Overview of experimental strategy}

For producing large condensates in optical dipole traps, it is
necessary to independently optimize both trap loading and
evaporative cooling. For initial loading of as many atoms as
possible, an optical trap with large volume is needed which, in
view of limited laser power, implies a shallow trapping potential.
For subsequent forced evaporative cooling, however, high densities
and fast elastic collisions require much tighter confinement.
These two requirements in general demand dynamical changes of the
trapping potential. A possible way to implement this is a spatial
compression of the optical trap using e.g.\ a zoom-lens system
\cite{Weiss2001}. Our approach is based on an alternative way
where a sequence of optical trapping schemes is used to provide
optimized loading together with optimized evaporative cooling.

We first use a shallow, large volume CO$_2$-laser trap as a
``reservoir'' for collecting the atoms before forced evaporative
cooling is implemented in a tighter trap. The reservoir trap can
be efficiently loaded with atoms that are precooled by
Raman-sideband cooling \cite{Kerman2000}. This approach allows
collection of atoms at moderate densities with little loss from
three-body collisions and with negligible heating from either
photon scattering or trap vibrations. It serves as a good starting
point for the final transfer into a tighter optical trap. The
tighter trap is adiabatically increased and adds a ``dimple'' to
the trapping potential of the reservoir. Collisional loading of
this dimple already yields a significant enhancement of the local
number and phase-space density \cite{StamperK1998}. After turning
off the reservoir trap excellent conditions for further forced
evaporative cooling are obtained.

\begin{figure*}[t]
\begin{center}
\epsfig{file=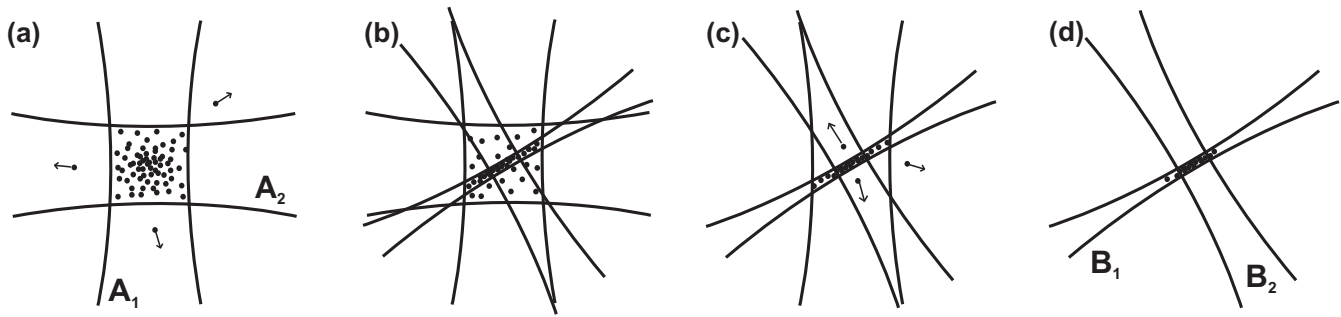,width=1.0\linewidth} \caption{ Illustration
of the various stages of trap loading and evaporative cooling as
seen from above. (a) Plain evaporation in a crossed CO$_2$-laser
trap generated by beams A$_1 $ and A$_2 $ at a scattering length
of $ a = 1215 \, a_0 $. (b) $ 1.5 $ s of ramping and collisional
loading into a crossed $ 1064 $-nm fibre laser trap generated by
beams B$_1 $ and B$_2 $ with a final scattering length  $ a = 210
\, a_0 $. (c) Forced evaporative cooling after switching off
CO$_2$-laser beam A$_2 $. The power of all remaining lasers is
ramped down, and the power in CO$_2$-laser beam A$_1$ is reduced
to zero. (d) Final configuration of the crossed $ 1064 $-nm trap.
Imaging is done in the horizontal plane at an angle of $
30^{\circ} $ with respect to the long axis of the cigar-shaped
atomic cloud. \label{crossedtrap} }
\end{center}
\end{figure*}

The different trap stages of optical trapping used in our
experiments are illustrated in Fig.~\ref{crossedtrap}. An overview
of the evolution of phase-space density and particle number for
the various trapping stages is shown in Fig.~\ref{psd}.

The use of relatively weak optical trapping necessitates the
implementation of magnetic ``levitation'' where a magnetic field
gradient along the vertical direction compensates for the
gravitational force. This levitation turns out to be very useful
in two ways: First, in the limit of very weak optical trapping
only one spin state is held in the trap. This assures perfect spin
polarization of the sample\footnote{This Stern-Gerlach separation
technique also allows for radio-frequency evaporation along the
vertical direction. Although one-dimensional, this type of
evaporation has been applied to produce ultracold Cs atoms for
studying three-body collisions \cite{Weber2003b}.}. Further,
efficient evaporation can be performed without the effect of
gravitational sag in the trap. The dc magnetic field offset
remains a free parameter for flexible tuning of the scattering
length.

\subsection{Laser cooling}

The initial collection and cooling of Cs atoms is achieved by
conventional techniques. In a stainless steel vacuum chamber
\cite{Weber2003c} atoms are loaded into a magneto-optical trap
(MOT) from a Zeeman slowed atomic beam with up to $ 3 \times 10^8
$ atoms after about $6$ s. The MOT is operated on the
$6^2$S$_{1/2}, F=4 $ to $6^2$P$_{3/2}, F'=5 $ transition. The
ultra-high vacuum of less than $ 1 \times 10^{-11} $ mbar gives
200 s for the $1/e$-lifetime of the MOT. The MOT light is derived
from a high power laser diode\footnote{SDL-5712-H1} referenced via
beat-lock to a grating-stabilized master diode laser. Standard
absorption imaging is used to determine particle numbers and
temperatures.

We compress the atomic cloud by ramping up the magnetic field
gradient in the MOT by a factor of 5 to $ 33 $ G/cm within $ 40 $
ms. Simultaneously we linearly change the detuning of the MOT
laser from around $ 10 $ MHz to $ 30 $ MHz. At the end of the
ramp, we switch off the MOT light and the magnetic field gradient.
To cool the compressed cloud, we then apply degenerate
Raman-sideband cooling \cite{Kerman2000} in an optical lattice to
further cool and to polarize the atoms in the desired $ F=3, m_F=3
$ state. We have adapted the technique as described in
\cite{Treutlei2001} to our setup. This cooling scheme is
particularly suited for polarizing atoms in the  $ F=3, m_F=3 $
state because this is a dark state for which photon scattering is
suppressed. Four laser beams derived from an injection locked
slave laser resonant with the $ F=4 $ to $ F'=4 $ transition
produce a three-dimensional
optical lattice, % $ 9.2 $ GHz detuned from the resonance for the
% atoms in the $ F=3 $ ground state. At the same time, their light
drive Raman-sideband transitions % between the different magnetic
% sublevels in the $ F=3 $ ground state. Also, the light serves to
and repump out of the $ F=4 $ ground state manifold. The total
power of all the four beams is 65 mW and their $ 1/e^2 $-beam
radii are about $ 1 $ mm. The oscillation frequency in the lattice
is on the order of $ 100 $ kHz. A small magnetic field offset of
several hundred mG is applied to induce the Raman-sideband
cooling. We succeed in polarizing $ 90 \% $ of the atoms. The
ensemble is then adiabatically released from the lattice after 6
ms of cooling time. If the atomic cloud is released into free
space, the temperature of the ensemble with up to $ 4 \times 10^7
$ atoms is about $ 0.7 \, \mu$K. For our typical atomic densities
this corresponds to a phase space density of $ 1 \times 10^{-3} $.

\subsection{Reservoir trap}\label{LevT}
We generate the large reservoir trap by horizontally crossing two
CO$_2$-laser beams A$_1 $ and A$_2 $ at right angles as shown in
Fig.~\ref{crossedtrap}(a). At the same time we apply a magnetic
gradient field in the vertical direction to levitate the atoms
against gravity. The delivered powers in laser beams A$_1 $ and
A$_2 $ are 90 W and 65 W, respectively. The light comes from two
separate, highly stable linearly polarized single-frequency
CO$_2$-lasers\footnote{Coherent-DEOS GEM-100L}. Switching of the
beams is done by external acousto-optical
modulators\footnote{Intraaction AGM-408BB1M} (AOMs). A$_1 $ is
downshifted in frequency by 40 MHz, whereas A$_2 $ is upshifted by
40 MHz to prevent any interference. To avoid mode-hops the cooling
water for the lasers needs to be stabilized to better than $ \pm
20 $ mK. Still, a slow mode drift changes the power of the lasers
by a few percent over the time scale of minutes. At the crossing
point the $ 1/e^2 $-beam radii of the two lasers are $ (605 \pm
35) \, \mu$m and $ (690 \pm 35) \, \mu$m.

The magnetic fields for levitation and for Feshbach tuning are
generated by two pairs of coils aligned with their axes parallel
to the vertical direction. One pair in anti-Helmholtz
configuration produces the vertical magnetic field gradient near $
31.3 $ G/cm to levitate the atoms in the $ F=3, m_F=3 $ state.
Another pair in Helmholtz configuration provides a variable bias
field $ B_0 $ of up to 200 G. The combined field results in a weak
outward directed force $ F(\rho) = m \alpha^2 \rho $ depending on
the horizontal distance $ \rho $ from the vertical symmetry axis.
For perfect levitation of our atoms the constant $ \alpha = g
\sqrt{m/(3 \mu_B B_0)} $ describes the curvature of the parabolic
anti-trapping potential. The levitation field thus slightly
reduces the trap depth along the horizontal direction. Here, $ m $
is the mass of Cs, $ g $ is the gravitational acceleration, and $
\mu_B $ is Bohr's magneton. At $ B_0 = 17 $ G we have $ \alpha = 2
\pi \times 3.4 $ Hz. The horizontal trap frequencies $
\omega_{x,y} $ are reduced according to $ \omega'_{x,y} =
\sqrt{\omega_{x,y}^2 - \alpha^2} $. This is usually a very small
effect for all but the lowest trap frequencies. Note that
levitation also affects the horizontal motion of free atoms after
the optical trap is shut off. The horizontal motion follows $
\rho(t) = \rho_0 \cosh{(\alpha t)} + \alpha^{-1} v_0 \sinh{(\alpha
t)} $ for initial position $ \rho_0 $ and initial velocity $ v_0
$. The vertical motion is not affected.

We excite vertical trap oscillations by briefly changing the
vertical magnetic field gradient and hence tilting the trap. For
exciting horizontal trap oscillations we shift the equilibrium
position of the atoms by adding a horizontal magnetic field
component. In both cases we monitor the center-of-mass oscillation
of the atomic cloud after $ 50 $ ms time-of-flight. The
geometrically averaged trap frequency $\bar{\nu}$ is calculated to
be $ (12.6 \pm 1.5) $ Hz which is in good agreement with the
experimental value of $ (13.2 \pm 0.2) $ Hz. Together with the
levitation and the magnetic bias field the lasers provide an
effective trap depth of about $k_B \times 7 \, \mu $K. This trap
depth is given by the weaker of the two CO$_2$-lasers as the atoms
can escape along the direction of the stronger beam.

For transfer of the precooled atoms into the reservoir trap, we
leave the light of the two CO$_2$-lasers on during the entire
pre-cooling phase. This is because the CO$_2$-lasers show strong
variations in beam pointing and beam shape as a function of
radio-frequency power to the AOMs. We have checked that the small
light shift introduced by the lasers does not affect the initial
loading and cooling efficiency. The reservoir trap is then
activated by ramping up the magnetic field and its gradient. The
$1/e$-rise time of the magnetic fields is limited to $ 1.5 $ ms
because of eddy currents in the stainless steel chamber. We
therefore do not expect the transfer to be fully adiabatic.

We find that the atoms are heated to about $ 2.2 \, \mu$K by the
transfer into the reservoir trap. A clear measurement on the
trapped sample is only possible after about $ 50 $ ms since the
system initially is not in thermal equilibrium and since the
untrapped atoms need to disappear from the field of view. We
largely attribute the heating to imperfect phase space matching.
In fact, the atomic cloud after Raman-sideband cooling to $ 0.7 \,
\mu$K has a $1/e$-radius of $ \sim 350 \, \mu$m. In comparison, an
equilibrium distribution in the reservoir trap at $ 0.7 \, \mu$K
would have a $1/e$-radius of $ \sim 100 \, \mu$m. Potential energy
is thus gained which is then turned into kinetic energy,
effectively heating the cloud of atoms. Subsequently, the hot
atoms evaporate out of the trap. For this phase of plain
evaporation we set the magnetic bias field to $ 73.5 $ G. The
scattering length is then $ 1215 \, a_{0}$. The temperature is
reduced to less than $ 1 \, \mu$K within $ 10 $ s. After this
time, we measure more than $ 4 \times 10^6 $ atoms, corresponding
to a peak phase space density of $ 2 \times 10^{-3} $.

\subsection{Dimple trap}

We proceed with loading of the dimple trap after $ 2 $ s of plain
evaporation in the reservoir trap. At this point the atom number
is $7.8 \times 10^{6}$ and the phase space density is $1.7 \times
10^{-3}$ (see Fig.~\ref{psd}). The dimple trap is generated by
horizontally intersecting one tightly focused laser beam B$_1 $
with $ 34$-$\mu$m waist and another less focused beam B$_2 $ with
$ 260$-$\mu$m waist at right angles, rotated by $ 30^{\circ} $ in
the horizontal plane with respect to the CO$_2$-laser beams as
shown in Fig.~\ref{crossedtrap}(d). This is different from our
earlier work \cite{Weber2003a} where we have used CO$_2$-laser
beam A$_2$ for axial confinement. We introduce the B$_2 $ beam
because some weak back reflections of the CO$_2$-laser beams led
to a slight undesirable corrugation of the optical potential. This
complicated the quantitative analysis of the BEC. Beams B$_1 $ and
B$_2 $ are derived from a broadband fiber laser\footnote{IPG Laser
PYL-10} at $1064$ nm. The powers in these beams are ramped up
within 1.5 s to a maximum power of $ 70 $ mW for B$_1 $ and $ 270
$ mW for B$_2 $. The trapping in the dimple is now briefly done by
all four laser beams with B$_1$ providing most of the radial and
A$_1$ most of the axial confinement. After switching off beam
A$_2$ we measure the radial and axial trap frequencies in the
dimple to $ (221.2 \pm 1.6) $ Hz and $ (14.2 \pm 0.1) $ Hz,
respectively. During the ramping up phase of B$_1 $ and B$_2 $ we
reduce the magnetic field offset to $ 23 $ G and thus the
scattering length to $ 300 \, a_0 $ in order to reduce losses from
three-body recombination \cite{Weber2003b}. The trap now contains
about $ 1.7 \times 10^6 $ atoms at a peak phase space density of
approximately $ 0.13 $.

\subsection{Forced evaporation towards BEC}

We start forced evaporative cooling by ramping down the power in
all three remaining beams. Simultaneously we remove the reservoir
by switching off the CO$_2$-laser A$_2$ that is not responsible
for axial confinement. To assure a well-defined ramp over a large
intensity range we control the light power of the near-infrared
beam B$_1$ by means of a logarithmic photodiode and a servo loop.
The power in CO$_2$-laser beam A$_1$ is ramped to zero within $
5.5 $ s so that B$_2$ at the end of evaporation exclusively
assures axial confinement. The change in beam pointing for A$_2$
does not affect the evaporation. For B$_1$ we approximately follow
an exponential ramp over 5.5 s. The power in beam B$_2$ is only
slightly reduced. The final power in B$_1$ and B$_2$ is $ 0.5 $ mW
and $ 220 $ mW. We find and optimize this ramp by extending the
ramp in discrete time steps of a few hundred milliseconds at the
beginning and up to one second towards the end of the ramp.

\begin{figure}[t]
\begin{center}
\epsfig{file=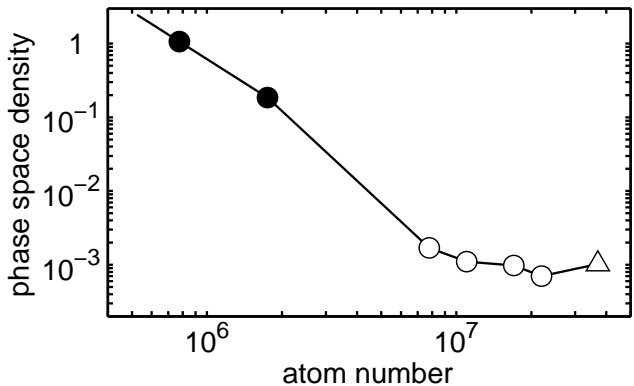,width=1.0\linewidth} \caption{Peak phase
space density as function of atom number. The path of evaporation
proceeds from right to left. The {\it triangle} shows the atomic
ensemble immediately after lattice cooling. The {\it open circles}
show the ensemble in the reservoir trap after 0.08, 0.22, 0.64,
and 2.0 s. The {\it filled circles} correspond to the sample in
dimple trap right after loading and after 1.5 s of evaporation.
The phase transition occurs after $ 2 $ s of forced evaporation
with $ \sim 5 \times 10^5 $ atoms left in the dimple trap.
\label{psd} }
\end{center}
\end{figure}

At each step we search for a maximum in evaporation efficiency
$\gamma = \log(D'/D) / \log(N/N')$ as a function of the trap depth
and scattering length \cite{Ketterl1996}. Here, $D$ and $D'$ are
the phase-space densities at the beginning and end of each step,
$N$ and $N'$ denote the respective particle numbers.
%Here, $\Delta \Phi$ denotes the gain in phase space density and
%$\Delta N$ the loss in particle number of each step.
Maximizing $ \gamma $ at each step results in an overall
optimization of the evaporation path. We find that a magnetic
field value of $21$ G with scattering length $ a = 210 \, a_0 $ is
optimal during the forced evaporation phase. As can be seen from
Fig.~\ref{psd} the efficiency $\gamma$ lies around 3 during the
forced evaporation ramp. We attribute this high efficiency to the
fact that atoms can escape the trap into almost all directions
because of the levitation field.

We observe the phase transition after $ 2 $ s of forced
evaporative cooling with about $ 5 \times 10^5 $ atoms at a
temperature of $ (200 \pm 10) $ nK. At this point the power in
beams B$_1$ and B$_2$ is $ 8.7 $ mW and $ 250 $ mW. The duration
of the ramp is relatively short. Our evaporation proceeds close to
the hydrodynamic regime. Thus, significant improvement of the
evaporation is not to be expected.

\begin{figure}[t]
\begin{center}
\epsfig{file=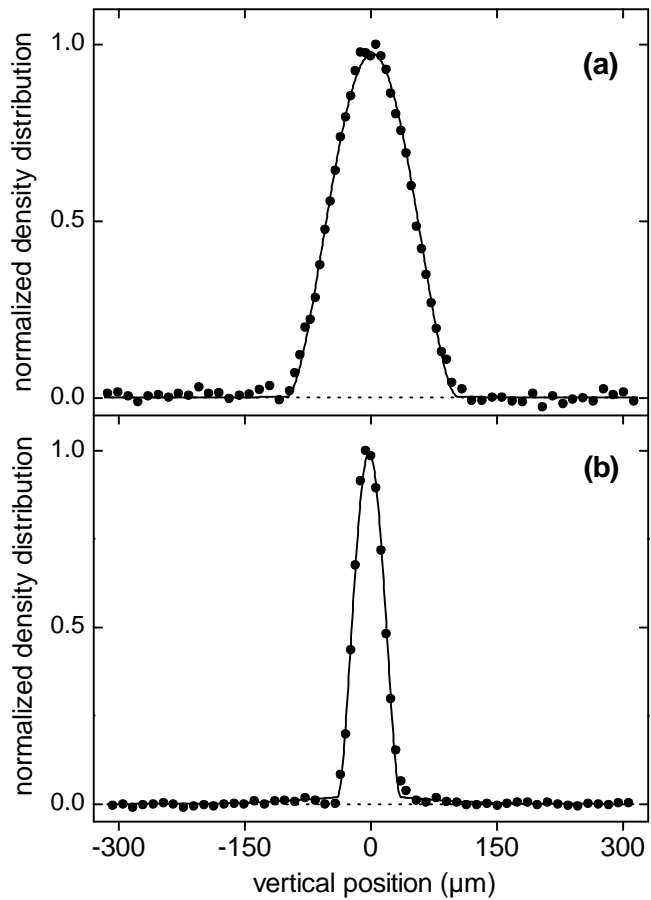,width=1.0\linewidth} \caption{ Vertical
density profiles of Cs condensates after $ 100 $ ms of free
expansion in the levitation field. The solid curves are fits to
the data for the Thomas-Fermi profiles which include possible
thermal components. For better distinction the baseline is dashed.
(a) Expansion with no change in scattering length. The total
number of atoms in the condensate is $ N = 1.1 \times 10^5 $. (b)
Expansion near zero scattering length under the same conditions
reveals a small thermal component with a temperature of about $10$
nK. \label{condensateprofile} }
\end{center}
\end{figure}

Further evaporation leaves a cigar-shaped condensate with the long
axis in the horizontal plane. In Fig.~\ref{condensateprofile} we
show vertical density profiles of expanding condensates. The
tunability of the scattering length allows us to explore different
regimes of expansion. For Fig.~\ref{condensateprofile}(a) we
expand the condensate at the creation scattering length of $ 210
\, a_0$. This is the usual type of self-similar expansion in which
the condensate in the Thomas-Fermi regime retains its parabolic
shape \cite{Pitaevsk2003}. For Fig.~\ref{condensateprofile}(b) we
step the scattering length to zero at the moment of release from
the trap. The mean-field interaction thus vanishes and the rate of
expansion is greatly reduced. This exposes a small thermal
component, for which a bimodal fit reveals a temperature of around
$ 10 $ nK. The critical temperature at these trapping conditions
is $ 24 $ nK, therefore the expected condensate fraction agrees
well with the measured value of $91 \%$. From the fit to the data
in Fig.~\ref{condensateprofile} we obtain that there are up to $
1.1 \times 10^5 $ atoms in the condensate with a $20 \%$
calibration error. The error does not come from the fit but from
the overall uncertainty in determining the atom number. Usually,
the error from absorption imaging alone is around $50 \%$, but we
can calibrate the atom number from measurements on the chemical
potential, see Sec. \ref{expEnergy}. For this particular
experiment we measure the final trap frequencies to $ (4.3 \pm
0.2) $ Hz and $ (21.1 \pm 0.2) $ Hz along the axial and radial
direction, respectively. We thus infer for the initial
Thomas-Fermi sizes $ R_r^{TF}= (8.7 \pm 0.3) \, \mu$m and $
R_a^{TF}= (42.5 \pm 1.2) \, \mu$m along the radial and axial
directions at a scattering length of $a=210 \, a_0$. The peak
density of the condensate is $n_0 = (2.1 \pm 0.1) \times 10^{13} $
cm$^{-3}$.

\section{Tunable quantum gas}
\label{Tunablequantumgas} We now test the tunability of the
condensate interaction. We first study the condensate expansion as
a function of scattering length \cite{Cornish2000} in two
different ways. We then specialize to the case when the
interaction energy is switched off and present improved results on
the ultra-slow expansion of the condensate in comparison with
earlier measurements in \cite{Weber2003a}. Finally, we excite
compression oscillations of the trapped condensate by suddenly
stepping the scattering length to a lower value.

\subsection{Expansion energy as a function of scattering
length}\label{expEnergy} We measure the release energy of the
condensate for slow and fast changes of the scattering length.
When we slowly vary the scattering length the wave function of the
trapped condensate can follow adiabatically and the condensate
remains in equilibrium. The release energy is proportional to the
chemical potential of the condensate at the given value of the
scattering length. The situation is different when we rapidly
switch the scattering length at the moment of condensate release.
The condensate then expands from a non-equilibrium state because
the wave function has not had time to adjust to the change in
interaction energy. This leads to strong changes for the rate of
condensate expansion in comparison to the expansion from
equilibrium.

\begin{figure}[t]
\begin{center}
\epsfig{file=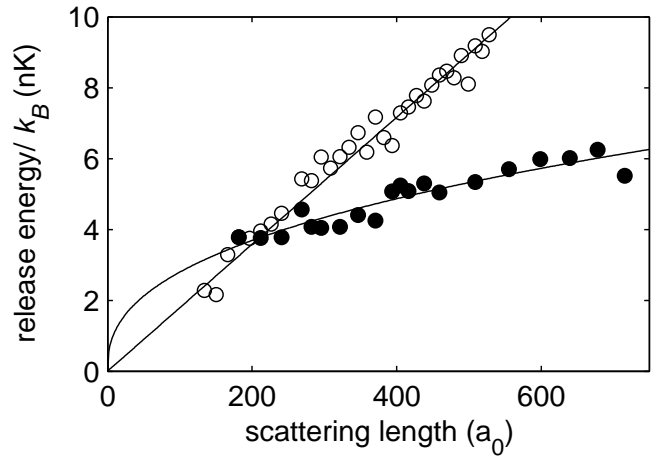,width=1.0\linewidth} \caption{Release energy
of the condensate as a function of scattering length $a$. The {\it
filled circles} represent experimental data for the case of
adiabatic ramping of a trapped condensate. The data, corresponding
to $2/7$ of the chemical potential at a given value of the
scattering length, is fit by $ C \, a^{2/5}$. The {\it open
circles} represent data for rapid switching at the moment of
condensate release. As discussed in the text, the straight line is
not a fit. It connects the origin with the fitted value of the
release energy at the creation scattering length. \label{chempot}}
\end{center}
\end{figure}

We first consider a condensate in the Thomas-Fermi regime for
which we adiabatically ramp the scattering length to a new value.
For such a condensate, the release energy $E_{rel}$ directly
corresponds to the chemical potential $\mu_{TF}$ through
$\frac{7}{2} E_{rel}=\mu_{TF}$ \cite{Pitaevsk2003}, which is given
by
\begin{equation}
\mu_{TF} = \frac{h \, \bar{\nu}}{2} \left( \frac{15 \, N }{a_{ho}}
\right)^{2/5} \, a^{2/5} . \label{equ1}
\end{equation}
Here, $ \bar{\nu} $ is the geometric average of the trap
frequencies, $ N $ is the particle number in the condensate, and
$a_{ho} = \sqrt{\hbar/(m \, 2 \pi \, \bar{\nu})}$ is the
oscillator length. For the experiment we produce a condensate with
$ N = 8.5 \times 10^4 $ atoms at a creation scattering length of
$a_c = 210 \, a_0$. We then slowly ramp the magnetic field to
values between 20 and 35 G, setting the scattering length to a
value between about $ 200 $ and $ 700 \, a_0 $. The slow ramping
excludes values below the Feshbach resonance at $ 19.9 $ and above
the one at $ 48.0 $ G because of strong loss\footnote{A
combination of slow ramping and quick jumping at the Feshbach
resonances would allow access to the full range of values for the
scattering length.}. The condensate is then released from the trap
and we measure the release energy. The results are shown in
Fig.~\ref{chempot}. Here we assume that the magnetic field
strength translates into scattering length according to
Fig.~\ref{Csscatteringlength}. The data is well fit by a function
of the form $ C \, a^{2/5}$ according to Eq.~(\ref{equ1}). From
the fit parameter $ C $ we can deduce an independent estimate of
the particle number $ N = (8.2 \pm 1.3) \times 10^4 $. The error
is dominated by the error in determining the trap frequencies.

For a sudden change of the scattering length the condensate wave
function has no time to react. For example, for an increase of the
scattering length the density distribution is too narrow in
comparison to the equilibrium density distribution at the new
value of the scattering length. The condensate thus expands more
rapidly than a condensate in equilibrium at this new value. Since
the mean-field interaction energy of the condensate scales
linearly with the scattering length for a given density profile
\cite{Pitaevsk2003}, we expect a linear behavior of the release
energy as a function of the final scattering length $a$. In
Fig.~\ref{chempot} we thus compare the data for the measured
release energy to a straight line $C \, a_c^{2/5} \, a/a_c $ given
by the origin and the fitted value of the release energy at the
creation scattering length $ a_c = 210 \, a_0$. We find good
agreement with the linear dependence.

\begin{figure}[t]
\begin{center}
\epsfig{file=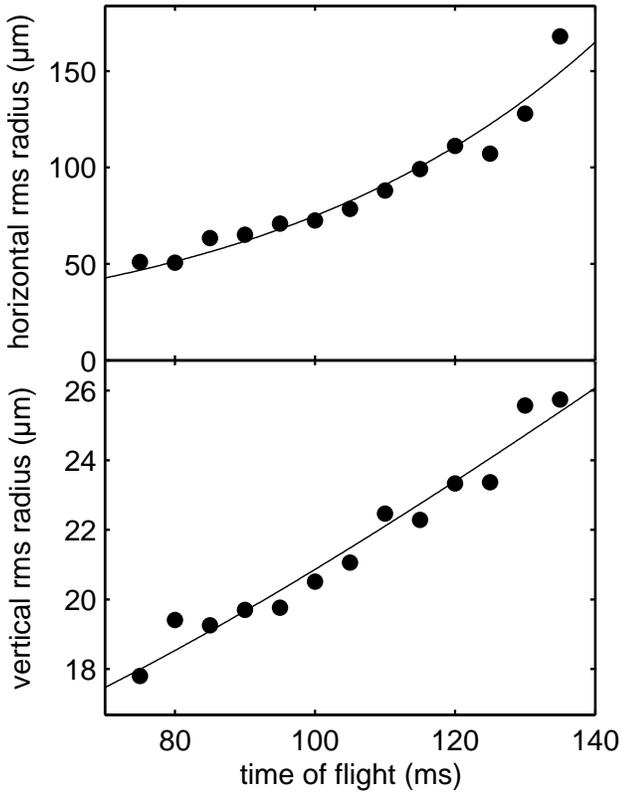,width=1.0\linewidth} \caption{Expansion of
the non-interacting condensate. The data points show the
horizontal (above) and vertical (below) rms radius of the BEC as a
function of expansion time near the zero crossing of scattering
length. Note the different scales. The fit to the residual
vertical expansion reveals a release energy of $k_B \times (51 \pm
3) $ pK. For the horizontal expansion the data is fit by $ A \,
\cosh{(\alpha \, t)}$ with $ \alpha = 2 \pi \times (3.20 \pm 0.23)
$ Hz. \label{frozen}}
\end{center}
\end{figure}

\subsection{Ultra-slow condensate expansion}
We now study the expansion of the condensate near the
zero-crossing of the scattering length. At the moment of
condensate release, we rapidly switch the magnetic field from the
creation field near $ 20 $ G to $ (17.17 \pm 0.05) $ G,
corresponding to $ a =(3.4 \pm 3.0) \, a_0$. The error in
determining the precise magnetic field at the position of
condensate requires that we choose a slightly positive value of
the scattering length to assure that no weakly attractive
interactions modify the condensate expansion. % The time constant
% for the switching of the magnetic field is again $ 1.5 $ ms.
The levitation field remains on, allowing for an extended
observation period because the atoms then do not fall under
gravity. Fig.~\ref{frozen} shows the vertical and horizontal
extent of a BEC with $ 1.2 \times 10^5 $ atoms as a function of
time after release from the trap. We only show the data after $75$
ms of expansion when the optical density of the atomic cloud is
sufficiently reduced to allow for reliable absorption imaging. The
horizontal expansion is dominated by the magnetic anti-trapping
potential which derives from the presence of the levitation field
and which magnifies the atomic cloud according to the cosine
hyperbolicus function, see Sec.~\ref{LevT}. The measured rate of
expansion $ 2 \pi \times (3.20 \pm 0.23) $ Hz agrees reasonably
well with the expected rate constant $ \alpha = 2 \pi \times 3.4 $
Hz. The vertical expansion corresponds to a release energy of $k_B
\times (51 \pm 3) $ pK. Note that this is much lower than the
kinetic energy of the ground state $ \hbar\omega_r/4 = k_B \times
253 $ pK given by a radial trap frequency of $ \omega_r = 2 \pi
\times \, 21.1 $ Hz. It is remarkable that the release energy is
less than the zero-point energy of the ground state. Since the
spatial extent of the condensate is much larger than the size of
the ground state wave function of the harmonic oscillator, the
momentum spread, limited by the uncertainty of the wave function
of the initial condensate, is lower than that of the ground state.

\subsection{Condensate oscillations}
\begin{figure}[t]
\begin{center}
\epsfig{file=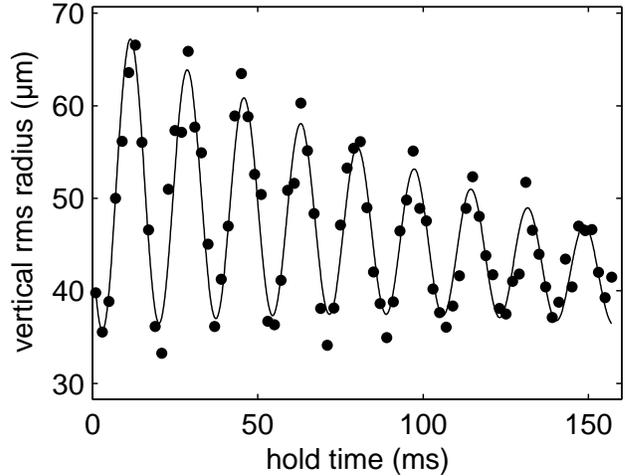,width=1.0\linewidth} \caption{Condensate
oscillations after rapid switching of the scattering length. The
{\it filled circles} show the vertical rms radius of an expanding
BEC with $ 7 \times 10^4 $ atoms after $ 80 $ ms of free expansion
as a function of hold time in the trap. The scattering length has
been switched rapidly from $ 363 \, a_0 $ to $ 25 \, a_0 $. The
{\it solid curve} is a fit to the data giving an oscillation
frequency of $ (58.3 \pm 0.2) $ Hz.
%and a $1/e$-decay time of $ 126 $ ms.
We independently measure the radial trap frequency to $ (28 \pm 1)
$ Hz. \label{wobble} }
\end{center}
\end{figure}

By rapidly ramping the scattering length it is possible to excite
oscillations of the condensate in the trap \cite{Kagan1997}. In
fact, in the limit of a cigar shaped condensate one expects radial
``compression'' or ``expansion oscillations'' at twice the trap
frequency. Compression oscillations can be seen in
Fig.~\ref{wobble} where we plot the vertical radius of the
released condensate as a function of hold time $ t_h $ in the
trap. To excite the oscillation we step the scattering length from
a value of $ a = 363 \, a_0 $ ($B=24.4 $ G) to $a = 25 \, a_0 $
($B=17.6 $ G) at time $ t_0 $. The condensate is then allowed to
oscillate in the trap for a variable hold time $ t_h $ at the
final value of the scattering length. We release the condensate at
time $ t_0 + t_h $ and take an image after $ 80 $ ms of free
expansion. We fit the data by a sinusoidal function.  The measured
compression oscillation frequency of $ (58.3 \pm 0.2) $ Hz agrees
well with twice the radial trap frequency of $ 2\times (28 \pm 1)
$ Hz at the given trapping power. To account for the damping we
have to introduce an exponential decay of the amplitude and of the
offset value. The damping of the amplitude has a time constant of
$ 126 $ ms. We have not yet identified the origin of this damping.
Possibly the BEC samples different trapping frequencies due to the
large amplitude of the oscillation, which would lead to an
apparent damping. Also, damping might be caused by the interaction
with a residual thermal cloud or by parametric processes
\cite{Chevy2002}.

\section{Conclusion}
\label{Sum} We have shown that essentially pure Cs condensates can
be produced with more than $10^5$ atoms. In our optical trap it is
possible to flexibly change the atomic scattering properties. The
atomic condensate can now be used as the starting point for
experiments where a tuning and ramping of the scattering
properties can be exploited. It will be interesting to study the
case of a non-interacting condensate at the zero-crossing of the
scattering length. Such a condensate might be used in atom
interferometers where one wishes to suppress any mean-field
effects \cite{Gupta2002}. On the other hand, tuning to large
values of the scattering length might allow the investigation of
effects beyond the mean-field approximation \cite{Pitaevsk2003}.
Also, modulation of the scattering length could be used as an
alternative tool to probe the excitation spectrum of the
condensate. Finally, ultracold Cs$_2$ molecules can be created by
ramping across one of the Feshbach resonances \cite{Herbig2003}
and the transition from an atomic to a molecular condensate could
then be studied.

\noindent {\bf Acknowledgements} This work is supported by the
Austrian ``Fonds zur F{\"o}rderung der wissenschaftlichen
Forschung'' (FWF) within SFB 15 (project part 16) and by the
European Union in the frame of the Cold Molecules TMR Network
under Contract No. HPRN-CT-2002-00290. M.M. is supported by DOC
[Doktorandenprogramm der \"{O}sterreichischen Akademie der
Wissenschaften]. C.C. is supported by a Lise-Meitner-Fellowship
from the FWF.

\bibliographystyle{prsty}
\bibliography{Lit}

\end{document}